\documentclass[proceedings, preprint]{rmaa}

\usepackage{paralist}

\usepackage{psfrag,color}

\SetYear{2008} \SetConfTitle{Magnetic Fields in the Universe II}

\title{Magnetic Visions:\\ Mapping Cosmic Magnetism with LOFAR and SKA}

\author{
  Rainer Beck\altaffilmark{1}}

\altaffiltext{1}{Max-Planck-Institut f\"ur Radioastronomie, Auf dem
H\"ugel 69, 53121 Bonn, Germany (rbeck@mpifr-bonn.mpg.de).}

\shortauthor{Beck}
\shorttitle{Magnetic Visions}

\listofauthors{R. Beck}
\indexauthor{Beck, R.}

\abstract{The origin of magnetic fields in the Universe is an open
problem in astrophysics and fundamental physics. ``Cosmic
Magnetism'' has been accepted as Key Science Project both for the
Low Frequency Array (LOFAR, under construction) and the planned
Square Kilometre Array (SKA). At low frequencies LOFAR and SKA will
allow to map the structure of weak magnetic fields in the outer
regions and halos of galaxies, in galaxy clusters and in the Milky
Way. High-resolution polarization observations at high frequencies
with the SKA will trace magnetic fields in the disks and central
regions of galaxies in unprecedented detail. All-sky surveys of
Faraday rotation measures (RM) towards polarized background sources
will be used to model the structure and strength of the magnetic
fields in the Milky Way, the interstellar medium of galaxies and the
intergalactic medium. The new method of ``RM Synthesis'', applied to
spectro-polarimetric data cubes, will separate RM components from
different distances and allow 3-D ``Faraday tomography''. Magnetic
fields in distant galaxies and clusters and in intergalactic
filaments will be searched for by deep imaging of weak synchrotron
emission and of RM towards background sources. This will open a new
era in the observation of cosmic magnetic fields.\\}

\resumen{}

\addkeyword{Techniques: polarimetric} \addkeyword{galaxies: halos}
\addkeyword{galaxies: magnetic fields} \addkeyword{galaxies: spiral}
\addkeyword{intergalactic medium} \addkeyword{radio continuum:
galaxies}

\begin{document}
\maketitle

\bigskip
\bigskip
\hspace{1cm}

\section{New-generation radio telescopes: LOFAR and SKA}
\label{sec:intro}

Magnetic fields of galaxies have been a topic of intensive
investigation since many years. However, many important questions,
like the origin and evolution of magnetic fields in galaxies,
especially their first occurrence in young galaxies, the presence of
magnetic fields in elliptical galaxies without active nucleus or the
small-scale structure of magnetic fields in the Milky Way remained
unanswered. The intracluster gas in galaxy clusters hosts magnetic
fields of considerable strength and coherence, but their origin is
not known. Finally, we would like to know whether the intergalactic
space is magnetized and whether outflows from galaxies or AGNs are
sufficient to maintain these fields.

\begin{figure}[!t]
\includegraphics[bb = 48 40 561 732,width=\columnwidth,clip=]{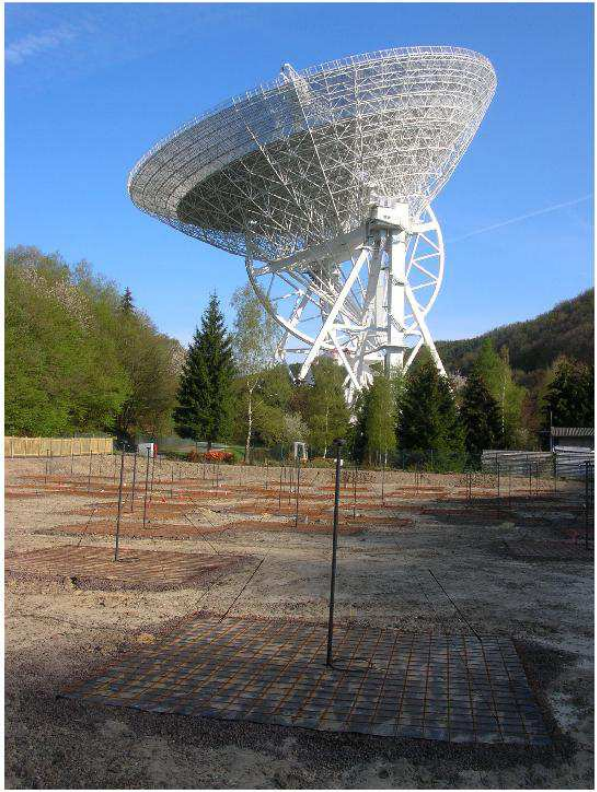}
\caption{The first international LOFAR station (front) with 96
low-band antennas (20--80~MHz) next to the Effelsberg 100~m
telescope (back) (Copyright: MPIfR Bonn).} \label{fig:lofar}
\end{figure}

Most of what we know about galactic magnetic fields comes through
the detection of radio waves. {\em Synchrotron emission} is related
to the total field strength in the sky plane, while its polarization
yields the orientation of the regular field in the sky plane and
also gives the field's degree of ordering. Incorporating {\em
Faraday rotation} provides information on the strength and direction
of the coherent field component along the line of sight.

The new generation of radio telescopes will open a new era of
magnetic field studies. {\em Cosmic Magnetism} is the topic of Key
Science Projects for LOFAR and the SKA \citep{gaensler04,beck07a}.
LOFAR ({\em Low Frequency Array}) is presently under construction
and will lead the way for a new generation of radio telescopes
consisting of a multitude of small and cheap antennas. It will work
in two frequency bands, 20--80~MHz and 110--240~MHz. The radio waves
are sampled digitally and the signals from the stations are
transmitted over large distances to a high-performance computing
facility, where the radio images are synthesized in real time. About
35 stations at distances up to about 100~km from the core will be
erected in 2008--2010 in the northern part of the Netherlands. The
first international station next to the Effelsberg 100~m telescope
is operating since 2007 (Fig.~\ref{fig:lofar}). Three more German
stations will follow in 2008--9. Further international stations are
funded in the UK, France and Sweden.

\begin{figure}[!t]
\includegraphics[bb = 48 40 561 388,width=\columnwidth,clip=]{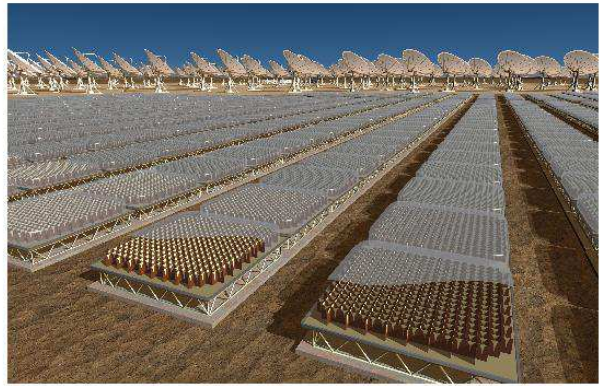}
\caption{SKA reference design: aperture array for low frequencies
and parabolic dishes for high frequencies (Copyright: SKA Programme
Development Office and XILOSTUDIOS).} \label{fig:ska}
\end{figure}

LOFAR can be considered as a pathfinder for a European participation
in the SKA ({\em Square Kilometre Array}), the next-generation
international radio telescope, which is envisaged for the years
beyond 2015. The SKA is planned to cover most of the radio window
accessible from the ground, from about 70~MHz to 35~GHz. The SKA
Reference Design (Fig.~\ref{fig:ska}) aims at two different
telescope arrays, one being a low-frequency digital telescope like
LOFAR.

Other SKA pathfinder telescopes under construction are ASKAP ({\em
Australian SKA Pathfinder}), consisting of two arrays operating
between 80~MHz and 2~GHz, ATA ({\em Allen Telescope Array}, USA,
1--10~GHz), LWA ({\em Long Wavelength Array}, USA, 10--88~MHz) and
MeerKat ({\em Karoo Array Telescope}, South Africa, 0.5--2.5~GHz).

Much of what LOFAR and SKA can contribute to our understanding of
magnetic fields will come from their {\em polarimetric
capabilities}. The crucial specifications are high polarization
purity and multichannel spectro-polarimetry. The former will allow
detection of the relatively low linearly polarized fractions from
most astrophysical sources, while the latter will enable accurate
measurements of Faraday rotation measures (RMs), intrinsic
polarization position angles and Zeeman splitting. The method of
{\em RM Synthesis}, based on multichannel spectro-polarimetry,
transforms the spectral data cube into a data cube of Faraday depth
\citep{brentjens05}. This allows to measure a large range of RM
values and to separate RM components from distinct regions along the
line of sight. If the structure of the medium along the line of
sight is not too complicated, this can be used for {\em Faraday
tomography}.

\bigskip

\section{Limitations of observations with present-day
radio telescopes}

\subsection{Extent of galactic magnetic fields in galaxies and galaxy clusters}
\label{sec:extent}

The observation of radio synchrotron emission from galaxies only
reveals magnetic fields illuminated by cosmic-ray electrons (CRE)
accelerated by supernova shock fronts in regions of strong star
formation. As the propagation of CRE is limited, radio images at
centimeter wavelengths (synchrotron + thermal) are mostly similar to
images of star-forming regions, as observed e.g. in the
far-infrared.

NGC~6946 is one of the best studied spiral galaxies in radio
continuum. Its synchrotron emission is observed until 25~kpc
distance from the center \citep{beck07b}, limited by the extent of
the star-forming disk, while neutral gas is detected until twice
larger radius \citep{boomsma05}. Magnetic fields may extend far away
from star-forming regions where the magneto-rotational instability
(MRI) \citep{sellwood99} serves as an energy source of turbulence
and field amplification.

The spectacular synchrotron ring of the Andromeda galaxy M~31 is the
result of an extended magnetic field of considerable regularity
illuminated by CRE generated in the relatively narrow ring of
star-formation at about 10~kpc from the center \citep{fletcher04}.
Faraday rotation measurements towards polarized background sources
showed that the regular field of M~31 exists also inside and outside
the ring \citep{han98}. However, the Faraday method could not be
applied to more galaxies because the number density of polarized
background sources is too small with the sensitivity of present-day
radio telescopes \citep{stepanov08}.

\begin{figure}[!t]
\includegraphics[bb = 27 25 516 493,width=\columnwidth,clip=]{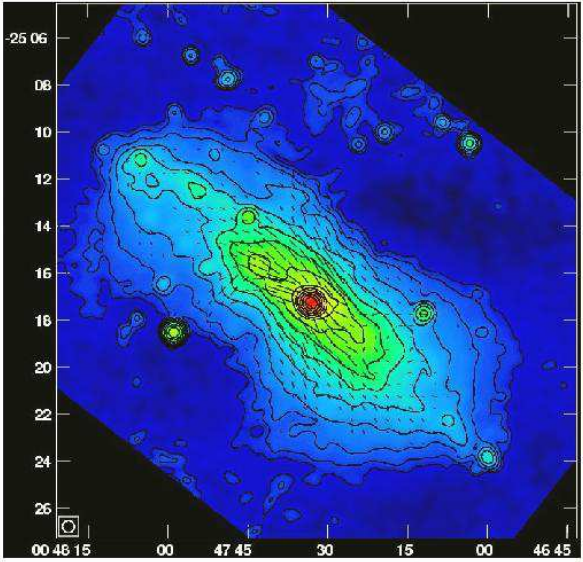}
\caption{Total radio emission (contours) and polarization
$B$--vectors of the almost edge-on spiral galaxy NGC~253, combined
from observations at 6~cm wavelength with the VLA and the Effelsberg
100-m telescope and smoothed to 30\arcsec\ resolution
\citep{heesen08} (Copyright: AIRUB Bochum).} \label{fig:n253}
\end{figure}

The observable size of radio halos around (almost) edge-on galaxies
generally increases with decreasing observation frequency, which
indicates that the extent is limited by energy losses of the CRE,
i.e. synchrotron, inverse Compton, bremsstrahlung and adiabatic
losses \citep{pohl90}, and by the advection velocity of the outflow
from the disk. NGC~253 is a prominent example (Fig.~\ref{fig:n253}).
The halo extent is smallest in the inner region where the magnetic
field is strongest and hence the synchrotron loss is highest
\citep{heesen08}. The X-shaped pattern of the B--vectors in the halo
of NGC~253 is typical for edge-on galaxies, possibly a signature of
a galactic wind with a radial component.

The vertical profile of radio halos of most edge-on galaxies
observed at 5~GHz can be described by an exponential decrease with
about 2~kpc vertical scaleheight \citep{krause04}. This corresponds
to a scaleheight of the magnetic field of about 8~kpc, assuming
equipartition between the energy densities of the field and the
total cosmic rays and a constant ratio of CR protons to electrons.
However, this ratio is expected to increase with increasing distance
from the disk due to the energy losses of the CRE, so that the
field's scaleheight is larger than 8~kpc. A similar argument refers
to the radial scalelengths of the radio disks.

\begin{figure}[!t]
\includegraphics[bb = 49 49 559 545,width=\columnwidth,clip=]{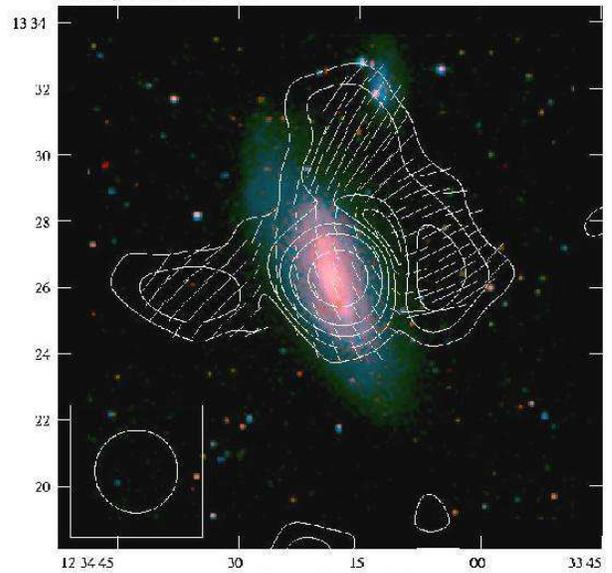}
\caption{Polarized radio emission (contours) and B-vectors of the
spiral galaxy NGC~4569 in the Virgo Cluster, observed at 6~cm
wavelength with the Effelsberg telescope \citep{chyzy06} (Copyright:
Cracow Observatory).} \label{fig:n4569}
\end{figure}

Interaction between galaxies or of a galaxy with the intergalactic
medium imprints unique signatures onto magnetic fields in galaxy
halos and thus onto the radio emission. The Virgo cluster is a
location of strong interaction effects. Highly asymmetric
distributions of the polarized emission shows that the magnetic
fields of several cluster spirals are strongly compressed on one
side of the galaxy \citep{vollmer07,wez07}. The lobes of the Virgo
spiral NGC~4569 reach out to at least 25~kpc from the disk and are
highly polarized (Fig.~\ref{fig:n4569}), probably a remainder of
interaction in the past.

The centers of some galaxy clusters host halos of radio synchrotron
emission (\citet{feretti04}, Brunetti, this volume). They are
probably created by turbulent wakes by the motion of galaxies
through the intracluster medium. The polarization of cluster halos
is low. Weak synchrotron emission was also detected around a group
of radio galaxies \citep{kronberg07}. Synchrotron radiation from
regions of large-scale shocks in clusters, where the intergalactic
magnetic fields are compressed and cooled pools of formerly highly
relativistic particles are re-accelerated, are called {\em relics}
(En{\ss}lin, this volume). Their radio emission is characterized by
strong polarization \citep{govoni05}. There may be large numbers of
exhausted radio sources, with "starved" AGN. Due to their steep
spectra, all of these structures are best detectable at low radio
frequencies.

Polarized radio emission is an excellent tracer of interactions in
the intergalactic and intracluster medium. As the decompression
timescale of the field is very long, it keeps memory of events in
the past. These are still observable if the lifetime of the
illuminating cosmic-ray electrons is sufficiently large.

To overcome the limitations by CRE lifetime, two observation methods
will be applied:

1. {\em Low frequencies:} CRE losses are smaller, and the extent of
galactic magnetic fields into intergalactic space can be traced. The
lifetime of CRE due to synchrotron losses increases with decreasing
frequency $\nu$ and decreasing total field strength $B$: $t_{syn} =
1.1~10^9$~yr $(\nu/GHz)^{-0.5}(B/\mu G)^{-1.5}$. In a $5~\mu$G field
the lifetime of electrons emitting in the LOFAR bands is
2-5~$10^8$~yr. In turbulent magnetic fields, cosmic rays propagate
by diffusion, with a diffusion speed equal to the Alfv\'en speed. In
the hot medium of galaxy or cluster halos (electron density $n_e
\simeq 10^{-3}$~cm$^{-3}$), the Alfv\'en speed is about 70~km/s
$(B/\mu$G). In field strengths above $3.25~\mu G \cdot (z+1)^2$
where synchrotron loss is stronger than loss due to Inverse Compton
with CMB photons, CRE radiating at 50~MHz can travel huge distances
of about 330~kpc $(B/\mu G)^{-0.5}$.

Polarized synchrotron emission traces ordered magnetic fields which
can be generated from turbulent fields by compressing or shearing
gas flows. Low frequencies will reveal such effects at larger
distances out to the intergalactic medium.

2. {\em Faraday rotation:} The halos of galaxies and galaxy clusters
contain hot gas which causes Faraday rotation of the polarized
emission from background sources if some fraction of the magnetic
field has a coherent component along the line of sight. As Faraday
rotation increases with the square of wavelength, low frequencies
are again preferable. In the outer halos of galaxies and in the
intracluster medium, we expect electron densities of $n_e \simeq
10^{-3}$~cm$^{-3}$ and field strengths of about 1~$\mu$G with 1~kpc
coherence length, causing a rotation measure of 0.7~rad/m$^2$ which
should be easily detectable with LOFAR and the low-frequency SKA.

The key platform on which to base the SKA's studies of cosmic
magnetism will be the {\em all-sky RM survey}\ at around 1~GHz which
can yield RMs towards about $10^7$ compact polarized extragalactic
sources \citep{gaensler04}. This data set will provide a grid of RMs
at a mean spacing of just $\simeq1-2'$ between extragalactic
sources.

\subsection{Dynamical effects of magnetic fields in galaxies}
\label{sec:dynamics}

\begin{figure}[!t]
\includegraphics[bb = 48 40 561 418,width=\columnwidth,clip=]{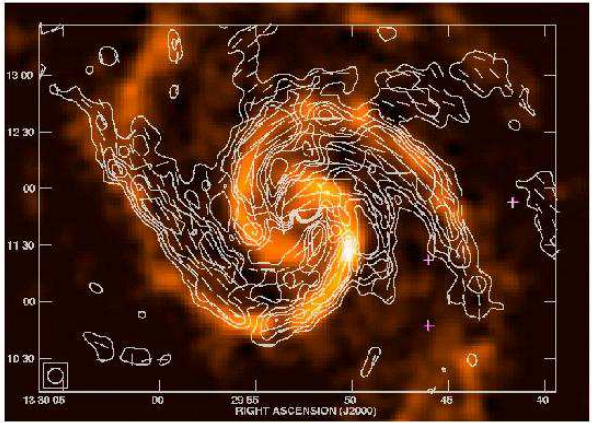}
\caption{Polarized radio emission (contours) and $B$--vectors of the
spiral galaxy M~51, combined from data at 6~cm wavelength from the
VLA and the Effelsberg 100-m telescopes and smoothed to 8\arcsec\
resolution \citep{fletcher08}. The background image shows the
integrated CO(1-0) line emission \citep{helfer03} (Copyright: MPIfR
Bonn).} \label{fig:m51}
\end{figure}

\begin{figure}[!t]
\includegraphics[bb = 49 57 561 634,width=\columnwidth,clip=]{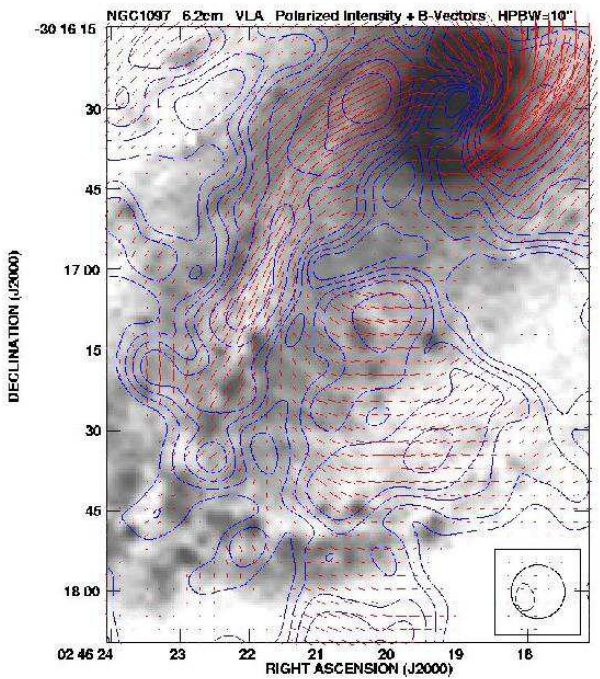}
\caption{Polarized radio emission (contours) and $B$--vectors of the
barred galaxy NGC~1097, smoothed to 10\arcsec\ resolution, observed
at 6~cm wavelength with the VLA \citep{beck+05}. The background
optical image is from Halton Arp (Copyright: MPIfR Bonn and Cerro
Tololo Observatory).} \label{fig:n1097}
\end{figure}

Surprisingly, large-scale shock fronts in the gas, found in spiral
arms and bars from spectral line observations, have only weak
counterparts in polarized synchrotron emission: The ordered field
avoids the shock. Striking examples are the spiral galaxy M~51
\citep{patrikeev06} (Fig.~\ref{fig:m51}) and the barred galaxy
NGC~1097 (Fig.~\ref{fig:n1097}). \citet{beck+05} argued that the
field is connected to the diffuse gas which has a large sound speed
and is not shocked. As the energy density of the field is comparable
to that of the diffuse gas, the gas velocity may be affected by the
field. The circumnuclear ring of NGC~1097 (Fig.~\ref{fig:n1097}, top
right) is another case of field--gas interaction: the field is
strong enough for an inward deflection of about one solar mass of
gas per year, sufficient to feed the active nucleus \citep{beck+05}.

Another indication for the dynamical importance of interstellar
magnetic fields comes from the comparison of energy densities in the
spiral galaxy NGC~6946. Beyond 5~kpc radius, the magnetic energy
density seems to be larger than the thermal energy density and the
kinetic energy density of turbulent cloud motions \citep{beck07b}.
``Super-equipartition'' fields may result from the {\em
magneto-rotational instability} (MRI) which transfers energy from
the shear of differential rotation into turbulent and magnetic
energy \citep{sellwood99}.

\citet{battaner00,battaner07} proposed that in the outermost parts
of spiral galaxies the magnetic field energy density may even reach
the level of global rotational gas motion and affect the rotation.
Fields in the outer regions can best be measured by Faraday rotation
of polarized emission from background sources with LOFAR and SKA.

If the magnetic field is dynamically important, dynamo and other MHD
models have to include the back-reaction onto the gas flow. Further
observational evidence should be provided by detailed comparisons
between field structures and gas velocity fields at spatial
resolutions of better than 100~pc. Present-day telescopes do not
provide sufficient sensitivity at such resolutions because the
signal from extended sources decreases with the beam area.

The best spatial resolution in synchrotron polarization achieved so
far in galaxies is about 10~pc in the LMC (Mao et al., in prep.).
100~pc can be reached in M~31, but only a few regions provide
signals significantly above the noise level. M~33 is faint in
polarization due to weakly ordered fields. IC~342 allows 200~pc
resolution and reveals bright polarized filaments, some of them
unrelated to gas structures \citep{beck05}. These data are too
sparse for a systematic study of the dynamical importance of
magnetic fields in galaxies. {\em Higher sensitivity} is needed to
allow studies at higher resolution.

\subsection{Magnetic field structure of the Milky Way}
\label{sec:milkyway}

The small-scale and large-scale structure of the magnetic field of
the Milky Way is topic of numerous investigations since several
decades. Nevertheless, no consistent picture could yet be
established.

Polarized synchrotron emission from the Milky Way and RM data from
polarized extragalactic sources were used by \citet{sun08} to model
the large-scale Galactic field. One large-scale reversal is required
about 1--2~kpc inside the solar radius. This confirmed the previous
analysis of \citet{brown07} that the local field runs clockwise
(seen from the northern Galactic pole) but reverses to
counter-clockwise towards the next inner spiral arm, the
Sagittarius-Carina arm, and is still counter-clockwise in the inner
molecular ring. \citet{sun08} claimed that the direction of the disk
field is continuous across the Galactic plane, but they needed a
second field component, called halo field, which is strongest at
1~kpc distance from the plane and reverses its direction across the
plane.

A model of the Galactic field based on 554 pulsar RM values
collected from several telescopes indicated reversals at several
Galactic radii, possibly between each spiral arm and the adjacent
interarm region \citep{han06}. The fields in the main inner arms
(Carina-Sagittarius, Scutum-Crux and Norma) run counterclockwise,
while the fields run clockwise in the interarm regions, in the solar
neighborhood and in the outer Perseus arm. On the other hand,
\citet{noutsos08}, based on independent measurements of 150 pulsar
RMs with the Parkes telescope, found safe evidence for two
reversals. The first is located about 1~kpc inside the solar radius,
between the local field (clockwise field) and the Carina arm
(counter-clockwise). The second reversal occurs between the Carina
and Crux arms at 2--4~kpc distance from the sun towards the Galactic
center. \citet{noutsos08} found the field in the Carina arm to
reverse from counter-clockwise to clockwise beyond about 4~kpc
distance from the sun, contrary to \citet{han06}. However,
sub-samples of RM values above and below the Galactic plane in
several regions revealed different field reversals, which calls for
caution with interpreting the data.

Though the spatial resolution with present-day radio telescopes in
external galaxies is much lower, typically a few 100~pc, large-scale
field reversals in the RM maps of diffuse polarized emission should
have been observed if they persist over at at least a few kpc and
their vertical extent is similar to that of the synchrotron-emitting
disk. However, only two large-scale reversals were found, one in the
flocculent galaxy NGC~4414 \citep{soida02} and one in the barred
galaxy NGC~1097 \citep{beck+05}. In both cases the line of reversal
runs at about constant azimuthal angle, different from the reversals
claimed for the Milky Way. Field reversals on smaller scales are
probably more frequent but also more difficult to observe in
external galaxies because the signal-to-noise ratios are quite low.
Only in the barred galaxy NGC~7479, where a jet serves as a bright
polarized background, several reversals on 1--2~kpc scale could be
detected in the foreground disk of the galaxy \citep{laine08}.
Either the Galactic reversals are not coherent over several kpc, or
they are restricted to a thin region near the Galactic plane, or the
Milky Way is special.

All Galactic field models so far assumed a constant pitch angle.
However, experience from external galaxies shows that the field's
pitch angle is not constant along the spiral arm and that the field
may even slide away from the spiral arms, as observed e.g. in M~51
\citep{patrikeev06}, but this cannot explain the discrepancy between
the Galactic and extragalactic observations.

A reliable model for the large-scale Galactic magnetic field needs a
much higher number of pulsar and extragalactic RM, hence much larger
sensitivity. The SKA ``Magnetism'' Key Science Project plans to
observe an all-sky RM grid (Sect.~\ref{sec:extent}) which should
contain about $10^4$ pulsar values with a mean spacing of $\simeq
30'$.

While the large-scale field is much more difficult to measure in the
Milky Way than in external galaxies, Galactic observations can trace
magnetic structures to much smaller scales. The 1.4~GHz all-sky
polarization map by \citet{wolleben06} reveals a wealth of details,
but suffers from Faraday depolarization near the Galactic plane.
Future polarization surveys will either observe at higher
frequencies (e.g. the Parkes S-PASS at 2.3~GHz, Carretti, this
volume) or observe at 1.4~GHz in the spectro-polarimetric mode with
a large number of channels (e.g. the Arecibo GALFACTS survey, Taylor
et al.). Multichannel data allow to apply the method of RM Synthesis
\citep{brentjens05}. Faraday depolarization can be reduced and
features at different distances can be separated. If the medium has
a relatively simple structure, {\em Faraday tomography} along
geometrical depth will be possible.

\subsection{Search for intergalactic magnetic fields}

The search for magnetic fields in the intergalactic medium (IGM) is
of fundamental importance for cosmology. All of ``empty'' space in
the Universe may be magnetized. Its role as the likely seed field
for galaxies and clusters and its possible relation to structure
formation in the early Universe, places considerable importance on
its discovery. A magnetic field already present at the epoch of
re-ionization or even at the recombination era might have affected
the processes occurring at these epochs \citep{subra06}. To date
there has been no detection of magnetic fields in the IGM; current
upper limits on the average strength of any such field suggest
$|B_{\rm IGM}| \le 10^{-8}-10^{-9}$~G \citep{kronberg94}.

Structure formation led to strong intergalactic shocks. Fields of
$B\simeq10^{-9}$--$10^{-8}$~G are expected along filaments of 10~Mpc
length with $n_e\simeq10^{-5}$~cm$^{-3}$ electron density
\citep{kronberg06}. This yields Faraday rotation measures of RM
$\simeq 0.1-1$~rad~m$^{-2}$. Their detection is a big challenge, but
possible. LOFAR has a realistic chance to measure intergalactic
magnetic fields {\em for the first time}. With LOFAR and SKA, the
synchrotron background emission from intergalactic shocks may also
become detectable below 500~MHz via their angular power spectrum
\citep{keshet04}.

If the all-pervading magnetic field will turn out to be even weaker,
it may still be identified through the planned all-sky RM grid with
the SKA (Sect.~\ref{sec:extent}). The correlation function of the RM
distribution provides the magnetic power spectrum as a function of
cosmic epoch \citep{blasi99}.

Primordial fields existing already in the recombination era would
induce Faraday rotation of the polarized CMB signals of the cosmic
microwave background (CMB) \citep{kosowsky96,kosowsky05} and
generate a characteristic peak in the CMB power spectrum at small
angular scales. The detection is challenging but possible with an
instrument of superb sensitivity like SKA.

\subsection{Origin of magnetic fields in galaxies} \label{sec:origin}

Radio observations of polarized synchrotron emission showed that the
large-scale field structures in almost all nearby galaxies are
spiral with pitch angles similar to those of the optical spiral arms
\citep{beck05}. Spiral fields were also detected in galaxies without
optical spiral structure and in circumnuclear gas rings. This
indicates the action of {\em dynamos} in rotating galaxies, where
the Coriolis force organizes turbulent gas motions, amplifying a
weak seed field and generating coherent field structures
\citep{beck96}. However, the physics of galactic dynamos is far from
being understood. The build-up of large-scale fields requires that
small-scale helicity is removed from the galaxy by outflows
\citep{sur06}. Even under favorable conditions, the timescale for
the growth of large-scale fields is several galactic rotation
periods.

{\em Primordial or protogalactic fields}, generated in the early
Universe or during galaxy formation, would be twisted by
differential rotation and destroyed by reconnection and diffusion
within a few rotation periods. Turbulent gas flows are needed to
maintain the field strength \citep{avillez05}. Compressing and
shearing gas flows, e.g. by spiral density waves, can generate
coherence on a scale of about 1~kpc \citep{otmia02}. To obtain
coherence on larger scales, the dynamo is the only available model.

The regular field structures obtained in models of the {\em
mean-field $\alpha\Omega$--dynamo}, driven by turbulent gas motions
and differential rotation, are described by modes of different
azimuthal symmetry in the disk. Such modes can be identified from
the pattern of polarization angles and Faraday rotation measures in
multi-wavelength radio observations of galaxy disks
\citep{krause90,elstner92} or from RM data of polarized background
sources \citep{stepanov08} and were found in the disks of a few
nearby spiral galaxies \citep{beck05}. However, in many galaxy disks
no clear patterns of Faraday rotation were found. Either many dynamo
modes are superimposed and cannot be distinguished with the limited
sensitivity and resolution of present-day telescopes, or no
large-scale dynamo modes exist and most of the ordered fields traced
by the polarization vectors are produced by compressing or shearing
gas flows. Again, {\em higher sensitivity} is needed because the
Fourier analysis of dynamo modes requires high-resolution RM data,
as planned by the SKA ``Magnetism'' Key Science Project
\citep{beck04}. The SKA will offer the chance to recognize these
patterns with help of Faraday rotation measures to at least 100~Mpc
distance \citep{stepanov08}.

Measurements of magnetic fields in distant galaxies (at redshifts
between $z\approx0.1$ and $z\approx2$) with SKA will provide direct
information on how magnetized structures evolve and amplify as
galaxies mature. The linearly polarized emission from galaxies at
these distances will often be too faint to detect directly. Faraday
rotation within these sources holds the key to studying their
magnetism. Available RM data of quasars at $z=2-3$ indicate that
their galaxy environment was already magnetized at $\mu$G levels
\citep{kronberg08}. Distant, but extended polarized sources like
radio galaxies also provide the ideal background illumination for
mapping Faraday rotation in galaxies along the same line of sight.

The mean-field $\alpha\Omega$--dynamo needs a few galactic rotations
or about $10^9$~yr to build up a coherent field of galactic scale
\citep{beck96}. The failure to detect coherent fields in distant
galaxies would indicate that the growth timescale is larger than the
galaxy age. If ``bisymmetric'' magnetic patterns turn out to
dominate, this would indicate that primordial or protogalactic
fields were twisted and amplified by differential rotation. If, on
the other hand, coherent fields are observed in young galaxies, this
would indicate that protogalactic fields were strong and possibly
already partly coherent.

At yet larger distances, with the sensitivity of the deepest SKA
fields, we expect to detect the synchrotron emission from
(unresolved) young galaxies and protogalaxies, since the turbulent
dynamo needs about $10^7$~yr to build up a galactic field of $\mu$G
strength \citep{brand05}. Unresolved galaxies will exhibit polarized
radio emission if their fields are ordered and their inclination is
larger than about 30$^{\circ}$ \citep{stil08}.

Radio telescopes operating at low frequencies (LOFAR, ASKAP-MWA and
the low-frequency SKA array) may also become very useful instruments
for field recognition or reconstruction with the help of background
RMs. The Faraday rotation angle increases with $RM \cdot \lambda^2$,
so that much smaller RM values can be detected at low frequencies.
However, the number density of polarized sources at low frequencies
is expected to be lower than at high frequencies due to Faraday
depolarization within the source and in the intervening medium.

\bigskip

\section{Conclusions}

Observing cosmic magnetism with present-day radio telescopes is
limited by low sensitivity and low resolution. Pushing these limits
requires large telescope arrays with large collecting areas, as
planned for LOFAR, SKA and the SKA pathfinders. Low frequencies and
spectro-polarimetry (RM Synthesis) can break the limits and open the
window to the magnetic Universe.

\bigskip



\begin{thebibliography}




\bibitem[{{de Avillez} \& {Breitschwerdt}(2005)}]{avillez05}
{de Avillez}, M.~A., \& {Breitschwerdt}, D.\ 2005, \aap, 436, 585

\bibitem[{{Battaner} \& {Florido}(2000)}]{battaner00} {Battaner}, E., \& {Florido}, E.\ 2000,
Fund. Cosmic Phys., 21, 1

\bibitem[{{Battaner} \& {Florido}(2007)}]{battaner07} {Battaner}, E., \& {Florido}, E.\ 2007,
Astron. Nachr., 328, 92

\bibitem[{{Beck}(2005)}]{beck05}
{Beck}, R.\ 2005, in: \textit{Cosmic Magnetic Fields.} Eds.\
R.~{Wielebinski} \& R.~{Beck}. (Berlin: Springer), p.~41


\bibitem[{{Beck}(2007a)}]{beck07a} {Beck}, R.\ 2007a, Adv. Radio Sci., 5, 399

\bibitem[{{Beck}(2007b)}]{beck07b} {Beck}, R.\ 2007b, \aap, 470, 539

\bibitem[{{Beck} \& {Gaensler}(2004)}]{beck04}
{Beck}, R., \& {Gaensler}, B.~M.\ 2004, New Astr. Rev., 48, 1289

\bibitem[{{Beck} {et~al.}(1996){Beck}, {Brandenburg}, {Moss}, {Shukurov}, \&
{Sokoloff}}]{beck96} {Beck}, R., {Brandenburg}, A., {Moss}, D.,
{Shukurov}, A., \& {Sokoloff}, D.\ 1996, \araa, 34, 155

\bibitem[{{Beck} {et~al.}(2005){Beck}, {Fletcher}, {Shukurov}, {Snodin},
  {Sokoloff}, {Ehle}, {Moss}, \& {Shoutenkov}}]{beck+05}
{Beck}, R., {Fletcher}, A., {Shukurov}, A., {et~al.}\ 2005, \aap,
444, 739

\bibitem[{{Blasi} {et~al.}(1999){Blasi}, {Burles}, \& {Olinto}}]{blasi99}
{Blasi}, P., {Burles}, S., \& {Olinto}, A.~V.\ 1999, \apj, 514, L79

\bibitem[{{Boomsma} {et~al.}(2005){Boomsma}, {Oosterloo}, {Fraternali},
{van der Hulst}, \& {Sancisi}}]{boomsma05} {Boomsma}, R.,
{Oosterloo}, T.~A., {Fraternali}, F., {van der Hulst}, J.~M., \&
{Sancisi}, R.\ 2005, in: \textit{Extraplanar Gas.} Ed.\ R.~{Braun}.
ASP Conf. Proc., 331, 247

\bibitem[{{Brandenburg} \& {Subramanian}(2005)}]{brand05}
{Brandenburg}, A., \& {Subramanian}, K.\ 2005, Phys. Rep., 417, 1

\bibitem[{{Brentjens} \& {de Bruyn}(2005)}]{brentjens05}
{Brentjens}, M.~A., \& {de Bruyn}, A.~G.\ 2005, \aap, 441, 1217

\bibitem[{{Brown} {et~al.}(2007){Brown}, {Haverkorn}, {Gaensler}, {et
al.}}]{brown07} {Brown}, J.~C., {Haverkorn}, M., {Gaensler}, B.~M.,
{et al.}\ 2007, \apj, 663, 258

\bibitem[{{Chy{\.z}y} {et~al.}(2006){Chy{\.z}y}, {Soida}, {Bomans}, {Vollmer},
{Balkowski}, {Beck}, \& {Urbanik}}]{chyzy06} {Chy{\.z}y}, K.~T.,
{Soida}, M., {Bomans}, D.~J., {et~al.}\ 2006, \aap, 447, 465

\bibitem[{{Elstner} {et~al.}(1992){Elstner}, {Meinel}, \& {Beck}}]{elstner92}
{Elstner}, D., {Meinel}, R., \& {Beck}, R.\ 1992, \aaps, 94, 587

\bibitem[{{Feretti} {et~al.}(2004){Feretti}, {Burigana}, \&
{En{\ss}lin}}]{feretti04} {Feretti}, L., {Burigana}, C., \&
{En{\ss}lin}, T.~A.\ 2004, New Astr. Rev., 48, 1137

\bibitem[{{Fletcher} {et~al.}(2004){Fletcher}, {Berkhuijsen}, {Beck},
{Shukurov}}]{fletcher04} {Fletcher}, A., {Berkhuijsen}, E.~M.,
{Beck}, R., \& {Shukurov}, A.\ 2004, \aap, 414, 53

\bibitem[{{Fletcher} {et~al.}(2008){Fletcher}, {Beck}, {Berkhuijsen},
{Horellou}, \& {Shukurov}}]{fletcher08} {Fletcher}, A., {Beck}, R.,
{Berkhuijsen}, E.~M., {Horellou}, C., \& {Shukurov}, A.\ 2008, \aap,
submitted

\bibitem[{{Gaensler} {et~al.}(2004){Gaensler}, {Beck}, \& {Feretti}}]
{gaensler04} {Gaensler}, B.~M., {Beck}, R., \& {Feretti}, L.\ 2004,
New Astr. Rev., 48, 1003

\bibitem[{{Govoni} {et~al.}(2005){Govoni}, {Murgia}, {Feretti}, {et
al.}}]{govoni05} {Govoni}, F., {Murgia}, M., {Feretti}, L., {et
al.}\ 2005, \aap, 430, L5

\bibitem[{{Han} {et~al.}(1998){Han}, {Beck}, \& {Berkhuijsen}}]{han98}
{Han}, J.~L., {Beck}, R., \& {Berkhuijsen}, E.~M.\ 1998, \aap, 335,
1117

\bibitem[{{Han} {et~al.}(2006){Han}, {Manchester}, {Lyne}, {Qiao},
\& {van Straten}}]{han06} {Han}, J.~L., {Manchester}, R.~N., {Lyne},
A.~G., {Qiao}, G.~J., \& {van Straten}, W.\ 2006, \apj, 642, 868


\bibitem[{{Heesen} {et~al.}(2008){Heesen}, {Beck}, {Krause}, \&
{Dettmar}}] {heesen08} {Heesen}, V., {Beck}, R., {Krause}, M., \&
{Dettmar}, R.-J.\ 2008, \aap, submitted


\bibitem[{{Helfer} {et~al.}(2003){Helfer}, {Thornley}, {Regan}, {et
al.}}]{helfer03} {Helfer}, T.~T., {Thornley}, M.~D., {Regan}, M.~W.,
et al.\ 2003, ApJS, 145, 259

\bibitem[{{Keshet} {et~al.}(2004){Keshet}, {Waxman}, \&
{Loeb}}]{keshet04} {Keshet}, U., {Waxman}, E., \& {Loeb}, A.\ 2004,
New Astr. Rev., 48, 1119

\bibitem[{{Kosowsky} \& {Loeb}(1996)}]{kosowsky96} {Kosowsky}, A.,
\& {Loeb}, A.\ 1996, \apj, 469, 1

\bibitem[{{Kosowsky} {et~al.}(2005)}]{kosowsky05} {Kosowsky}, A.,
{Kahniashvili}, T., {Lavrelashvili}, G., \& {Ratra}, B.\ 2005, Phys.
Rev. D, 71, 043006

\bibitem[{{Krause}(1990)}]{krause90}
{Krause}, M.\ 1990, in: \textit{Galactic and Intergalactic Magnetic
Fields.}\ Eds. R.~{Beck}, R.~{Wielebinski}, \& P.~P. {Kronberg}.
(Dordrecht: Kluwer), p.~187

\bibitem[{{Krause}(2004)}]{krause04}
{Krause}, M.\ 2004, in: \textit{The Magnetized Interstellar Medium.}
Eds. B.~{Uyaniker}, W.~{Reich}, \& R.~{Wielebinski}. (Katlenburg:
Copernicus), p.~173

\bibitem[{{Kronberg}(1994)}]{kronberg94} {Kronberg}, P.~P.\ 1994,
Rep. Prog. Phys., 57, 325

\bibitem[{{Kronberg}(2006)}]{kronberg06} {Kronberg}, P.~P.\ 2006,
Astr. Nachr., 327, 517

\bibitem[{{Kronberg} {et~al.}(2007){Kronberg}, {Kothes}, {Salter},
\& {Perillat}}]{kronberg07} {Kronberg}, P.~P., {Kothes}, R.,
{Salter}, C.~J., \& {Perillat}, P.\ 2007, \apj, 659, 267

\bibitem[{{Kronberg} {et~al.}(2008){Kronberg}, {Bernet}, {Miniati},
{et~al.}}]{kronberg08} {Kronberg}, P.~P., {Bernet}, M.~L.,
{Miniati}, F., {et~al.}\ 2008, \apj, 676, 70

\bibitem[{{Laine} \& {Beck}(2008)}]{laine08} {Laine}, S., \& {Beck},
R.\ 2008, \apj, 673, 128

\bibitem[{{Noutsos} {et~al.}(2008){Noutsos}, {Johnston}, {Kramer},
\& {Karastergiou}}]{noutsos08} {Noutsos}, A., {Johnston}, S.,
{Kramer}, M., \& {Karastergiou}, A.\ 2008, MNRAS, in press,
arXiv:0803.0677

\bibitem[{{Otmianowska-Mazur} {et~al.}(2002){Otmianowska-Mazur}, {Elstner},
{Soida} \& {Urbanik}}]{otmia02} {Otmianowska-Mazur}, K., {Elstner},
D., {Soida}, M., \& {Urbanik}, M.\ 2002, \aap, 384, 48

\bibitem[{{Patrikeev} {et~al.}(2006){Patrikeev}, {Fletcher}, {Stepanov},
{Beck}, {Berkhuijsen}, {Frick}, \& {Horellou}}]{patrikeev06}
{Patrikeev}, I., {Fletcher}, A., {Stepanov}, R., {et~al.}\ 2006,
\aap, 458, 441

\bibitem[{{Pohl} \& {Schlickeiser}(1990)}]{pohl90}
{Pohl}, M., \& {Schlickeiser}, R.\ 1990, \aap, 234, 147

\bibitem[{{Sellwood} \& {Balbus}(1999)}]{sellwood99}
{Sellwood}, J.~A., \& {Balbus}, S.~A.\ 1999, \apj, 511, 660

\bibitem[{{Soida} {et~al.}(2002){Soida}, {Beck}, {Urbanik}, \& {Braine}}]
{soida02} {Soida}, M., {Beck}, R., {Urbanik}, M., \& {Braine}, J.\
2002, \aap, 394, 47

\bibitem[{{Stepanov} {et~al.}(2008){Stepanov}, {Arshakian}, {Beck},
{Frick}, \& {Krause}}]{stepanov08} {Stepanov}, R., {Arshakian},
T.~G., {Beck}, R., {Frick}, P., \& {Krause}, M.\ 2008, \aap, 480,
45, 2008

\bibitem[{{Stil} {et~al.}(2008){Stil}, {Krause}, {Beck}, \& {Taylor}}]
{stil08} {Stil}, J.~M., {Krause}, M., {Beck}, R., \& {Taylor},
A.~R.\ 2008, in press, arXiv:0802.1374

\bibitem[{{Subramanian}(2006)}]{subra06} {Subramanian}, K.\ 2006,
Astr. Nachr., 327, 403

\bibitem[{{Sun} {et~al.}(2008){Sun}, {Reich}, {Waelkens}, \&
{En{\ss}lin}}]{sun08} {Sun}, X.~H., {Reich}, W., {Waelkens}, A., \&
{En{\ss}lin}, T.~A.\ 2008, \aap, 477, 573

\bibitem[{{Sur} {et~al.}(2006){Sur}, {Shukurov}, \&
{Subramanian}}]{sur06} {Sur}, S., {Shukurov}, A., \& {Subramanian},
K.\ 2006, MNRAS, 377, 874

\bibitem[{{Vollmer} {et~al.}(2007){Vollmer}, {Soida}, {Beck},
{Urbanik}, {Chy{\.z}y}, {Otmianowska-Mazur}, {Kenney}, \& {van
Gorkom}}]{vollmer07} {Vollmer}, B., {Soida}, M., {Beck}, R.,
{et~al.}\ 2007, \aap, 464, L37

\bibitem[{{We{\.z}gowiec} {et~al.}(2007){We{\.z}gowiec}, {Urbanik},
{Vollmer}, {Beck}, {Chy{\.z}y}, {Soida}, \& {Balkowski}}]{wez07}
{We{\.z}gowiec}, M., {Urbanik}, M., {Vollmer}, B., {et~al.}\ 2007,
\aap, 471, 93

\bibitem[{{Wolleben} {et~al.}(2006){Wolleben}, {Landecker}, {Reich},
\& {Wielebinski}}]{wolleben06} {Wolleben}, M., {Landecker}, T.~L.,
{Reich}, W., \& {Wielebinski}, R.\ 2006, \aap, 448, 411

\end{thebibliography}
\end{document}